\documentclass[12pt]{article}

\usepackage{amsmath,bbm,latexsym}
\usepackage{epsfig}
\usepackage{epstopdf}

\newcommand{\be}{\begin{equation}}
\newcommand{\ee}{\end{equation}}
\newcommand{\ba}{\begin{eqnarray}}
\newcommand{\ea}{\end{eqnarray}}
\newcommand{\no}{\nonumber\\}

\begin{document}

\title{
\LARGE A soft origin for CKM-type CP violation
}

\author{P.M.~Ferreira,$^{(1,2)}$\thanks{E-mail: ferreira@cii.fc.ul.pt} \
L.~Lavoura,$^{(3)}$\thanks{E-mail: balio@cftp.ist.utl.pt} \
and Jo\~{a}o P.~Silva$^{(1,3)}$\thanks{E-mail: jpsilva@cftp.ist.utl.pt}
\\*[3mm]\small $^{(1)}$ Instituto Superior de Engenharia de Lisboa \\
\small 1959-007 Lisboa, Portugal
\\*[2mm]
\small $^{(2)}$ Centro de F\'\i sica Te\'orica e Computacional,
Universidade de Lisboa \\
\small 1649-003 Lisboa, Portugal
\\*[2mm]
\small $^{(3)}$ Centro de F\'\i sica Te\'orica de Part\'\i culas,
Instituto Superior T\'ecnico \\
\small 1049-001 Lisboa, Portugal
}

\date{\today}

\maketitle

\vspace{-13cm}
\hfill \normalsize CFTP/11-005
\vspace{13cm}

\begin{abstract}
We present a two-Higgs-doublet model, with a $\mathbbm{Z}_3$ symmetry,
in which CP violation originates solely in a soft (dimension-2) coupling
in the scalar potential, and reveals itself
solely in the CKM (quark mixing) matrix.
In particular,
in the mass basis the Yukawa interactions of the neutral scalars are all real.
The model has only eleven parameters to fit the six quark masses
and the four independent CKM-matrix observables.
We find regions of parameter space in which
the flavour-changing neutral couplings are so suppressed
that they allow the scalars to be no heavier than a few hundred GeV.
\end{abstract}

\section{Introduction and notation}

One of the conceptually simplest extensions of the Standard Model (SM)
of the electroweak interactions
consists in allowing for $n_H > 1$ gauge-SU(2) ``Higgs'' doublets.
In such multi-Higgs-doublet models (MHDMs)
CP violation may occur in various places:
in the quark mixing matrix (CKM matrix) just as in the SM,
in the Yukawa couplings of the scalars to the quarks,\footnote{In this paper
we neglect the lepton sector.}
in the mixing of the scalars (in particular, scalar--pseudoscalar mixing),
or in the self-interactions (cubic and quartic interactions) among the scalars.
Unfortunately,
MHDMs in general lead to the existence of flavour-changing neutral currents
(FCNC),\footnote{More precisely,
quark-flavour-changing Yukawa interactions of the neutral scalars.}
which are severely restricted by the experimental data.

The simplest MHDMs are,
of course,
two-Higgs-doublet models (2HDMs) \cite{lee},
which have lately been the object
of intense scrutiny \cite{2HDMs}.
The Yukawa interactions of the quarks in the 2HDM are written
\be
\mathcal{L}_\mathrm{Yuk} = - \bar Q_L \sum_{k=1}^2
\left( \phi_k \Gamma_k n_R + \tilde \phi_k \Delta_k p_R \right)
+ \mathrm{H.c.},
\label{yuk1}
\ee
where $\phi_{1,2}$ are the two scalar gauge-SU(2) doublets,
$\tilde \phi_k \equiv i \tau_2 \phi_k^\ast$
for $k = 1, 2$,
$\Gamma_k$ and $\Delta_k$ are (in general, complex)
$3 \times 3$ matrices in flavour space,
and $Q_L$,
$n_R$,
and $p_R$ denote the 3-vectors (in flavour space)
of quark left-handed doublets,
right-handed charge $-1/3$ quarks,
and right-handed charge $+2/3$ quarks,
respectively.
In order for the U(1) gauge group of electromagnetism to be preserved,
the Higgs doublets are assumed to have vacuum expectation values (VEVs)
of the form
\be
\left\langle 0 \left| \phi_k \right| 0 \right\rangle =
\left( \begin{array}{c} 0 \\ v_k e^{i \theta_k} \end{array} \right),
\quad
\left\langle 0 \left| \tilde \phi_k \right| 0 \right\rangle =
\left( \begin{array}{c} v_k e^{- i \theta_k} \\ 0 \end{array} \right),
\ee
with real and non-negative $v_k$.
The quark mass matrices are then
\ba
M_n &=& \sum_{k=1}^2 v_k e^{i \theta_k} \Gamma_k,
\label{Mn}\\
M_p &=& \sum_{k=1}^2 v_k e^{- i \theta_k} \Delta_k.
\label{Mp}
\ea
These are bi-diagonalized as usual by unitary matrices $U_{L,R}^{n,p}$,
\ba
& & {U_L^n}^\dagger M_n U_R^n = M_d
= \mathrm{diag} \left( m_d, m_s, m_b \right),
\\
& & {U_L^p}^\dagger M_p U_R^p = M_u
= \mathrm{diag} \left( m_u, m_c, m_t \right),
\ea
and the CKM matrix is $V = {U_L^p}^\dagger U_L^n$.
The quantity $v = \sqrt{v_1^2 + v_2^2}
= \left( 2 \sqrt{2} G_F \right)^{-1/2} \approx 174\, \mathrm{GeV}$
is responsible for the masses of the $W^\pm$ and $Z^0$ gauge bosons.
It is convenient to use the `Higgs basis',
\ba
H_1 &=& \left. \left(
v_1 e^{- i \theta_1} \phi_1 + v_2 e^{- i \theta_2} \phi_2
\right) \right/ \! v
\\
&=& \left( \begin{array}{c}
G^+ \\ v + \left( h + i G^0 \right) \left/ \sqrt{2} \right.
\end{array} \right),
\\
H_2 &=& \left. \left(
v_2 e^{- i \theta_1} \phi_1 - v_1 e^{- i \theta_2} \phi_2
\right) \right/ \! v
\\
&=& \left( \begin{array}{c}
C^+ \\ \left( H + i A \right) \left/ \sqrt{2} \right.
\end{array} \right),
\ea
in which only $H_1$ has VEV,
which is precisely $v$.
The fields $G^+$ and $G^0$ are the would-be Goldstone bosons.
The field $C^+$ is a physical charged scalar.
The neutral fields $h$,
$H$,
and $A$ in general mix to form the three physical neutral scalars of the 2HDM.
We define the matrices
\ba
N_n &=& v_2 e^{i \theta_1} \Gamma_1 - v_1 e^{i \theta_2} \Gamma_2,
\\
N_p &=& v_2 e^{- i \theta_1} \Delta_1 - v_1 e^{- i \theta_2} \Delta_2,
\ea
and
\ba
N_d &=& {U_L^n}^\dagger N_n U_R^n,
\\
N_u &=& {U_L^p}^\dagger N_p U_R^p.
\ea
Equation~(\ref{yuk1}) then becomes
\ba
\mathcal{L}_\mathrm{Yuk} &=& - \bar d_L M_d d_R - \bar u_L M_u u_R
\no & &
- \frac{h}{\sqrt{2} v} \left( \bar d_L M_d d_R + \bar u_L M_u u_R \right)
- \frac{H}{\sqrt{2} v} \left( \bar d_L N_d d_R + \bar u_L N_u u_R \right)
\no & &
- \frac{i G^0}{\sqrt{2} v} \left( \bar d_L M_d d_R - \bar u_L M_u u_R \right)
- \frac{i A}{\sqrt{2} v} \left( \bar d_L N_d d_R - \bar u_L N_u u_R \right)
\no & &
+ \frac{G^+}{v}\,
\bar u \left( M_u V \gamma_L - V M_d \gamma_R \right) d
+ \frac{C^+}{v}\,
\bar u \left( N_u^\dagger V \gamma_L - V N_d \gamma_R \right) d
\no & &
+ \mathrm{H.c.},
\label{yuk2}
\ea
where $d$ and $u$ denote the column vectors in flavour space
of the charge $-1/3$ and charge $+2/3$ quarks,
respectively,
in the mass basis,
and $\gamma_{L,R}$ are the chirality projection matrices in Dirac space.
Since the matrices $N_d$ and $N_u$
are not necessarily diagonal,
the terms $\bar d_L N_d d_R$ and $\bar u_L N_u u_R$ in general include
potentially problematic FCNC.

We spot in the Yukawa interactions of equation~(\ref{yuk2})
three possible manifestations of CP violation:
\begin{description}
\item The CKM matrix $V$ may contain a complex phase,
just as in the SM.
\item The matrices $N_d$ and $N_u$ may be complex.
\item The scalars $h$ and $H$ may mix with the pseudoscalar
$A$.\footnote{If this mixing exists,
\textit{i.e.}\ if the three physical neutral scalars are mixtures
of all three $h$,
$H$,
and $A$,
then $A$ is not a physical particle and it does not make sense
to separate the physical neutral scalars
into two scalars and one pseudoscalar.}
\end{description}
One further manifestation of CP violation may occur in the cubic
and quartic interactions among the scalars.
It is the purpose of this paper to present a 2HDM
with an additional symmetry
such that
only the first one of the above four manifestations of CP violation occurs;
namely,
the matrices $N_d$ and $N_u$ are real,
the scalars do not mix with the pseudoscalar,
and the cubic and quartic interactions among the (neutral and charged) scalars
respect CP invariance.
Additionally,
our model shows that the FCNC may be quite suppressed
even when all the scalars have relatively low
(less than $1\, \mathrm{TeV}$) masses.

\section{The model: Yukawa couplings}
\label{sec:yuk}

Our model is a 2HDM supplemented by a particular $\mathbbm{Z}_3$ symmetry
and by
the usual CP symmetry.
Let $\omega = \exp{\left( 2 i \pi / 3 \right)}$.
Then,
under the $\mathbbm{Z}_3$ symmetry,
the following matter fields transform as
\be
\begin{array}{l}
\phi_2 \to \omega^2 \phi_2,
\\
Q_{L1} \to \omega^2 Q_{L1}, \quad Q_{L2} \to \omega Q_{L2},
\\
n_{R3} \to \omega n_{R3}, \quad p_{R3} \to \omega p_{R3},
\end{array}
\ee
and all other fields remain invariant.
This symmetry
forces the Yukawa-coupling matrices to have the following form
\cite{FerSilva}:
\[
\Gamma_1, \Delta_1 \sim \left( \begin{array}{ccc}
0 & 0 & 0 \\ 0 & 0 & \times \\ \times & \times & 0
\end{array} \right),
\quad
\Gamma_2 \sim \left( \begin{array}{ccc}
\times & \times & 0 \\ 0 & 0 & 0 \\ 0 & 0 & \times
\end{array} \right),
\quad
\Delta_2 \sim \left( \begin{array}{ccc}
0 & 0 & \times \\ \times & \times & 0 \\ 0 & 0 & 0
\end{array} \right),
\]
where the symbol $\times$ denotes
a
non-zero matrix entry.
The standard CP symmetry forces all those non-zero entries
of the Yukawa-coupling matrices to be real.
Therefore,
the mass matrices end up being
\ba
M_n &=& e^{i \theta_1} \left( \begin{array}{ccc}
e^{i \theta} x & 0 & 0 \\ 0 & 0 & a \\ b & c & e^{i \theta} y
\end{array} \right),
\label{Mmat1}
\\
M_p &=& e^{- i \theta_1} \left( \begin{array}{ccc}
0 & 0 & e^{- i \theta} a' \\
e^{- i \theta} b' & e^{- i \theta} c' & x' \\
y' & 0 & 0
\end{array} \right),
\label{Mmat2}
\ea
where $\theta = \theta_2 - \theta_1$ and $a$,
$b$, $c$, $x$, ..., and $y'$ are real.
In equation~(\ref{Mmat1})
we have already assumed a rotation
between $n_{R1}$ and $n_{R2}$ which renders zero
the $\left( 1, 2 \right)$ entry of $\Gamma_2$;
in the same way, in equation~(\ref{Mmat2})
a rotation between $p_{R1}$ and $p_{R2}$
has been used to make $\left( \Delta_1 \right)_{32} = 0$.
The matrices parameterizing the Yukawa couplings of $H_2$ are
\ba
N_n &=& e^{i \theta_1} \left( \begin{array}{ccc}
- e^{i \theta} x/r & 0 & 0 \\
0 & 0 & r a \\
r b & r c & - e^{i \theta} y/r
\end{array} \right),
\label{Nn}
\\
N_p &=& e^{- i \theta_1} \left( \begin{array}{ccc}
0 & 0 & - e^{- i \theta} a'/r \\
- e^{- i \theta} b'/r & - e^{- i \theta} c'/r & r x' \\
r y' & 0 & 0
\end{array} \right),
\label{Np}
\ea
where $r = v_2 / v_1$.

Let
\ba
{O_L^n}^T \left( \begin{array}{ccc}
|x| & 0 & 0 \\ 0 & 0 & |a| \\ |b| & |c| & |y|
\end{array} \right) O_R^n &=& M_d,
\label{Md} \\
{O_L^p}^T \left( \begin{array}{ccc}
0 & 0 & |a'| \\ |b'| & |c'| & |x'| \\ |y'| & 0 & 0
\end{array} \right) O_R^p &=& M_u,
\label{Mu}
\ea
where $O_{L,R}^{n,p}$ are real orthogonal matrices.
It is then clear that
\ba
{U_L^n}^\dagger &=&
{O_L^n}^T\, \mathrm{diag}\, \left(
1, \frac{abxy}{|abxy|}\, e^{2 i \theta},
\frac{bx}{|bx|}\, e^{i \theta} \right),
\\
U_R^n &=& e^{- i \theta_1}\, \mathrm{diag} \left(
\frac{x}{|x|}\, e^{- i \theta},\,
\frac{bcx}{|bcx|}\, e^{- i \theta},\,
\frac{bxy}{|bxy|}\, e^{- 2 i \theta}
\right)\, O_R^n,
\\
{U_L^p}^\dagger &=&
{O_L^p}^T\, \mathrm{diag}\, \left(
1, \frac{a'x'}{|a'x'|}\, e^{- i \theta},
\frac{a'b'x'y'}{|a'b'x'y'|}\, e^{- 2 i \theta} \right),
\\
U_R^p &=& e^{i \theta_1}\, \mathrm{diag} \left(
\frac{a'b'x'}{|a'b'x'|}\, e^{2 i \theta},\,
\frac{a'c'x'}{|a'c'x'|}\, e^{2 i \theta},\,
\frac{a'}{|a'|}\, e^{i \theta}
\right)\, O_R^p.
\ea
The CKM matrix is
\be
V =
{O_L^p}^T\, \mathrm{diag}
\left( 1, e^{i \alpha}, \pm e^{i \alpha} \right) O_L^n,
\label{V}
\ee
where
\be
e^{i \alpha} = \frac{a'x'aybx}{|a'x'aybx|}\, e^{- 3 i \theta},
\quad
\pm e^{i \alpha} = \frac{a'x'b'y'bx}{|a'x'b'y'bx|}\, e^{- 3 i \theta}.
\label{CKM_phase}
\ee
One sees that
\emph{the complexity of the CKM matrix
originates exclusively from the phase $3 \theta$},
which is the only phase with physical consequences in our model.
One easily finds the matrices parametrizing the non-diagonal Yukawa couplings:
\ba
N_d &=& {O_L^n}^T \left( \begin{array}{ccc}
- |x|/r & 0 & 0 \\ 0 & 0 & r |a| \\ r |b| & r |c| & - |y| / r
\end{array} \right) O_R^n,
\label{Nd} \\
N_u &=& {O_L^p}^T \left( \begin{array}{ccc}
0 & 0 & - |a'|/r \\ - |b'| / r & - |c'| / r & r |x'| \\ r |y'| & 0 & 0
\end{array} \right) O_R^p.
\label{Nu}
\ea
These matrices are real.
Thus,
\emph{in our model there is no CP violation from the FCNC matrices}.

\section{The model: scalar potential}

The scalar potential of our model is
\ba
V &=& V_\mathrm{sym} + V_\mathrm{SB},
\\
V_\mathrm{sym} &=&
\mu_1 \phi_1^\dagger \phi_1
+ \mu_2 \phi_2^\dagger \phi_2
\no & &
+ \frac{\lambda_1}{2} \left( \phi_1^\dagger \phi_1 \right)^2
+ \frac{\lambda_2}{2} \left( \phi_2^\dagger \phi_2 \right)^2
+ \lambda_3\, \phi_1^\dagger \phi_1\, \phi_2^\dagger \phi_2
+ \lambda_4\, \phi_1^\dagger \phi_2\, \phi_2^\dagger \phi_1,
\hspace*{5mm} \\
V_\mathrm{SB} &=&
- \left| \mu_3 \right| \left(
e^{- i \vartheta} \phi_1^\dagger \phi_2
+ e^{i \vartheta} \phi_2^\dagger \phi_1
\label{VSB}
\right),
\ea
where $V_\mathrm{sym}$
respects the $\mathbbm{Z}_3$ and CP symmetries of the model
while $V_\mathrm{SB}$ breaks both those symmetries,
but only softly.
The soft-breaking term is unique
and is as general as possible.
Note that $V_\mathrm{sym}$ coincides with
the Peccei--Quinn potential \cite{PQ}.
The minimization of the potential
leads to the vacuum phase $\theta$
being equal to the phase $\vartheta$ in $V_\mathrm{SB}$.
Thus,
in our model \emph{the origin of CP violation lies exclusively
in a soft term in the scalar potential}.~\footnote{
Note that ours is a model with soft CP breaking---the
Lagrangian does not enjoy CP symmetry because of the presence
of the $\mu_3$ term.
This is distinct from a model~\cite{branco} in which
spontaneous CP violation is achieved through the addition to the
Lagrangian of a soft (dimension-2) term which breaks
some other internal symmetry but does not break CP.
Spontaneous CP violation usually leads to CP violation
in the scalar sector, in particular through scalar--pseudoscalar mixing.
However, recently a model was found~\cite{CP3} in which there is spontaneous CP
violation but the scalar sector still preserves CP.
}

The equations for vacuum stability
read,
besides $\theta = \vartheta$,
\ba
\mu_1 &=& \left| \mu_3 \right| \frac{v_2}{v_1}
- \lambda_1 v_1^2 - \left( \lambda_3 + \lambda_4 \right) v_2^2,
\\
\mu_2 &=& \left| \mu_3 \right| \frac{v_1}{v_2}
- \lambda_2 v_2^2 - \left( \lambda_3 + \lambda_4 \right) v_1^2.
\ea

If we define
\be
m_A^2 = \frac{\left| \mu_3 \right| v^2}{v_1 v_2},
\quad
m_C^2 = m_A^2 - \lambda_4 v^2,
\ee
then we easily find that the part of $V$ which is bilinear in the fields is
\ba
V_\mathrm{bilinear} &=&
\frac{m_A^2}{2} \left( A^2 + H^2 \right) + m_C^2 C^- C^+
\no & &
+ \frac{\lambda_1 v_1^4 + \lambda_2 v_2^4
+ 2 \left( \lambda_3 + \lambda_4 \right) v_1^2 v_2^2}{v^2}\, h^2
\no & &
+ \left[ \lambda_1 + \lambda_2
- 2 \left( \lambda_3 + \lambda_4 \right) \right] \frac{v_1^2 v_2^2}{v^2}\, H^2
\no & &
+ 2 v_1 v_2\, \frac{\lambda_1 v_1^2 - \lambda_2 v_2^2
+ \left( \lambda_3 + \lambda_4 \right)
\left( v_2^2 - v_1^2 \right)}{v^2}\, h H.
\ea
One sees that $A$ does not mix with $h$ and $H$.
\emph{In our model there is no scalar--pseudoscalar mixing}.

Moreover,
\emph{in our model there is no CP violation in the self-interactions
of the scalars}.
This follows from the fact that
in a general 2HDM there is only one gauge-invariant
vacuum phase---$\theta$---and
in our specific 2HDM there are only two terms in the scalar potential---those
with coefficient $\left| \mu_3 \right|
\exp{\left( \pm i \vartheta \right)}$---which are sensitive to that phase.
The vacuum phase adjusts in such a way as to offset the phase
of those terms in the scalar potential
so that the final potential has no phase at all.

\section{The fit: procedure}

\subsection{First stage}

As seen in equations~(\ref{Md}) and~(\ref{Mu}),
the six quark masses depend only on ten parameters:
$|a|$, $|b|$, $|c|$, $|x|$, $|y|$,
$|a'|$, $|b'|$, $|c'|$, $|x'|$, and $|y'|$.
Then,
from equation~(\ref{V}),
the CKM matrix $V$,
which contains four independent observables,
depends on one additional parameter,
the phase
$\theta$.\footnote{The CKM
matrix additionally depends on the signs of
$a' x' b x a y$ and of $a' x' b x b' y'$,
as seen in equation~\eqref{CKM_phase}.}
One thus has to fit ten observables
by means of eleven parameters.\footnote{Even when
the number of parameters is larger
than the number of observables to be fitted,
obtaining a good fit is not always possible.
The fact that our model passes this test is
interesting in itself.}

We have assumed throughout that
the contributions to quark decays
from tree-level diagrams with intermediate scalars
are much smaller than the contributions
from diagrams with intermediate $W^\pm$.
We thus assume that the SM extractions of $|V_{us}|$,
$|V_{cb}|$,
and $|V_{ub}|$ still hold in our model.
These three CKM-matrix elements
and the quark masses
are allowed to take any value within their
Particle Data Group (PDG) allowed ranges \cite{PDG}.
In our fits $|V_{td}|$ is left free,
but we have found that,
once the various experimental constraints to be discussed below are included,
a good fit is obtained only when $|V_{td}|$ lies roughly
in the SM-allowed range.

We then proceed to analyze the FCNC of our model.
These are governed by the matrices $N_d$ and $N_u$
in equations~(\ref{Nd}) and~(\ref{Nu}),
respectively.
Those matrices involve the extra parameter $r=v_2/v_1$.

In our analysis of the FCNC, we consider only their contributions
to the mixing in the neutral-meson--antimeson systems $K$,
$B_d$,
$B_s$,
and $D$.
The relevant quantity is the off-diagonal matrix element $M_{12}$
connecting each meson to the corresponding antimeson.
That matrix element receives contributions both from an SM box diagram
and a tree-level diagram involving the FCNC.
We denote the latter by NP (for ``New Physics'') and write
\be
M_{12} = M_{12}^\mathrm{SM} + M_{12}^\mathrm{NP}.
\label{M12Total}
\ee

In order to shorten our text
we shall follow the notation in the textbook~\cite{BLS}
and freely use its equations with the prefix BLS.
For the $K$ system,
$M_{12}^\mathrm{SM}$ and the quantities relevant for its determination
can be found in equations~(BLS-17.14),
(BLS-17.16),
(BLS-B.15),
(BLS-B.16),
and (BLS-13.50);
the expressions for the other neutral-meson systems are obtained
by straightforward modifications of the quarks and mesons involved.
The quark masses and CKM-matrix elements
utilized in the calculation of $M_{12}$
are those produced by each of our fits;
in addition,
we use some other quantities shown in Appendix~A.

The calculation of $M_{12}^\mathrm{NP}$
is in equation (BLS-22.76).\footnote{That equation contains a
sign mistake in the hadronic matrix elements in the
vacuum-insertion approximation,
which we have corrected.}
This calculation requires
equations (BLS-22.29),
(BLS-22.33),
and (BLS-22.73).
Since there is no scalar--pseudoscalar mixing in our model,
one has
\be
M_{12}^\mathrm{NP} = M_{12}^A + M_{12}^{Hh},
\ee
where
\begin{description}
\item $M_{12}^A$ originates in the tree-level exchange
of the pseudoscalar
(parity-odd) $A$;
\item $M_{12}^{Hh}$ originates in the tree-level exchange
of the two physical parity-even scalars $S_1$ and $S_2$,
with masses $m_1$ and $m_2$,
respectively.
\end{description}
The scalars are mixtures of $H$ and $h$ through
\be
\left( \begin{array}{c} H \\ h \end{array} \right)
= \left( \begin{array}{cc}
\cos{\psi} & \sin{\psi} \\ - \sin{\psi} & \cos{\psi}
\end{array} \right)
\left( \begin{array}{c} S_1 \\ S_2 \end{array} \right).
\ee
One defines an effective mass $m_\mathrm{eff}$ in the scalar sector:
\be
\frac{1}{m_\mathrm{eff}^2}
= \frac{\sin^2{\psi}}{m_1^2} + \frac{\cos^2{\psi}}{m_2^2}.
\label{eq:meff}
\ee
One then has,
for the $K$ system,
\ba
M_{12}^A &=&
\frac{f_K^2 m_K}{192 v^2}\, \frac{1}{m_A^2} \left\{
- \left[ 1 + \frac{m_K^2}{\left( m_s + m_d \right)^2} \right]
\left[ \left( N_d \right)_{21}^\ast - \left( N_d \right)_{12} \right]^2
\right.
\no & &
\left.
+ \left[ 1 + \frac{11 m_K^2}{\left( m_s + m_d \right)^2} \right]
\left[ \left( N_d \right)_{21}^\ast + \left( N_d \right)_{12} \right]^2
\right\},
\label{M12A}
\\
M_{12}^{Hh} &=&
\frac{f_K^2 m_K}{192 v^2}\, \frac{1}{m_\mathrm{eff}^2} \left\{
\left[ 1 + \frac{m_K^2}{\left( m_s + m_d \right)^2} \right]
\left[ \left( N_d \right)_{21}^\ast + \left( N_d \right)_{12} \right]^2
\right.
\no & &
\left.
- \left[ 1 + \frac{11 m_K^2}{\left( m_s + m_d \right)^2} \right]
\left[ \left( N_d \right)_{21}^\ast - \left( N_d \right)_{12} \right]^2
\right\}.
\label{M12Hh}
\ea
Both $m_K$ and $f_K$ are given in Appendix~A.
In equations~(\ref{M12A}) and (\ref{M12Hh}),
we should note that the matrix $N_d$ is
real in our model,
therefore both $M_{12}^A$ and $M_{12}^{Hh}$
are real.

In the $K$ system,
we use $M_{12}$ to fit
\ba
\Delta m_K &=& 2 \left| M_{12} \right|,
\label{DeltamK}
\\
e^{- i \pi / 4}\, \epsilon_K &=&
- \frac{\mathrm{Im} \left( M_{12} \lambda_u^2 \right)}
{\sqrt{2}\, \Delta m_K \left| \lambda_u \right|^2},
\label{epsilonK}
\ea
where $\lambda_u = V_{us}^\ast V_{ud}$.
In the $K$ system there are
important long-distance contributions to $M_{12}$,
which we do not know how to compute precisely.
Therefore,
in that system we use for $M_{12}^\mathrm{SM}$
only the short-distance box diagrams,
but allow $\Delta m_K$
calculated by using equations~(\ref{M12Total}) and~(\ref{DeltamK})
to be in between one half and twice the experimental value.

In the $B_d$ and $B_s$ systems\footnote{In those systems
we use for $M_{12}^\mathrm{SM}$
a simplified expression involving only the exchange
of top quarks in the box diagram.}
we fit $\Delta m_{B_d}$ and $\Delta m_{B_s}$
by using a formula analogous to equation~(\ref{DeltamK}).

There are uncertainties in the ``bag parameters''
used in $M_{12}^\mathrm{SM}$.
In $M_{12}^\mathrm{NP}$,
we use the vacuum-insertion approximation
to calculate the values of the hadronic matrix elements,
and do not allow for corrections
to the matrix elements provided by that approximation.
In order to allow for these theoretical uncertainties,
we let our results for $\epsilon_K$,
$\Delta m_{Bd}$,
and $\Delta m_{Bs}$ differ from the experimental values by
at most
10\%.

We fit two more quantities,
$\sin{\left( 2 \beta \right)}$ and $\sin{\left( 2 \alpha \right)}$.
These are computed in the following way.
For the $K$,
$B_d$,
and $B_s$ decays we define\footnote{This definition
uses the sign conventions in~\cite{BLS}.
Many authors use instead $q \rightarrow - q$.}
\be
\frac{q}{p} = \frac{M_{12}^\ast}{\left| M_{12} \right|}.
\label{qoverp}
\ee
CP violation in $B_d \rightarrow \psi K_S$ is determined by
\be
\lambda_{\psi K_S} =
\left( \frac{q}{p} \right)_{B_d}
\frac{V_{cb} V_{cs}^\ast}{V_{cb}^\ast V_{cs}}
\left( \frac{p}{q} \right)_K.
\label{lambda-psiKS}
\ee
By using equations~(BLS-28.24),
(BLS-30.34),
and~(BLS-30.35),
we know that in the SM
$\lambda_{\psi K_S} = \exp{\left( - 2 i \beta \right)}$,
where $\beta$ is a certain phase
of the CKM matrix.\footnote{The phase $\epsilon^\prime$
in equation~(BLS-28.24) is known to be tiny.}
We therefore use
\be
\sin{\left( 2 \beta \right)} = - \mathrm{Im}\, \lambda_{\psi K_S}
\ee
and compare
our $- \mathrm{Im}\, \lambda_{\psi K_S}$
to the current experimental value of $\sin{\left( 2 \beta \right)}$.
In this way we constrain the NP contributions to $M_{12}$
in both the $B_d$ and $K$ systems,
through equations~\eqref{qoverp} and~\eqref{lambda-psiKS}.

An isospin analysis of the decays $B_d \rightarrow \pi \pi$
may be used,
together with the analysis of $B_d \rightarrow \rho \pi$
and $B_d \rightarrow \rho \rho$,
to extract
\be
\lambda_{\pi \pi}
= \left( \frac{q}{p} \right)_{B_d}
\frac{V_{ub} V_{ud}^\ast}{V_{ub}^\ast V_{ud}}.
\label{lambda-pipi}
\ee
Using equations~(BLS-28.24) and~(BLS-30.35),
we see that in the SM
$\lambda_{\pi \pi}
= - \exp{\left( 2 i \alpha \right)}$.
We thus use
\be
\sin{\left( 2 \alpha \right)} =
- \mathrm{Im}\, \lambda_{\pi \pi}
\ee
together with the current experimental value
of $\sin{\left( 2 \alpha \right)}$
to constrain $M_{12}^\mathrm{NP}$ in the $B_d$ system.

To summarize,
we work with 14 parameters:
$a$, $b$, $c$, $x$, $y$,
$a'$, $b'$, $c'$, $x'$, $y'$,
$\theta$, $r = v_2/v_1$, $m_A$, and $m_\mathrm{eff}$.
With those 14 parameters we strive to fit 15 observables:
$m_u$, $m_c$, $m_t$,
$m_d$, $m_s$, $m_b$,
$\left| V_{us} \right|$, $\left| V_{cb} \right|$,
$\left| V_{ub} \right|$,
$\Delta m_K$, $\epsilon_K$, $\Delta m_{B_d}$, $\Delta m_{B_s}$,
$\sin{\left( 2 \beta \right)}$, and $\sin{\left( 2 \alpha \right)}$.
We found that the fit is possible and,
indeed,
we have found a large variety of input parameters,
\textit{i.e.}\ of points in parameter space,
which are able to satisfy the criteria of the fit.

\subsection{Second stage}

Each one  of the fits found in the previous subsection
is \textit{a posteriori}\ passed through a filter,
to ensure that\footnote{We find that all the
fit points
which have passed through the filter
actually have $\left| V_{td} \right|$ in the SM range.}
\begin{description}
\item the Yukawa couplings are perturbative;
\item the quantity $\sin{\left( 2 \beta \right)}$
computed from the decays $B_d \rightarrow D^+ D^-$ is correct;
\item the angle $\gamma$ lies in the allowed range;
\item $\Delta m_D$ is not too large.
\end{description}
We next explain each of these four points.

From equation~(\ref{Mmat1})
we see that the Yukawa-coupling matrix $\Gamma_1$
has matrix elements $a/v_1$,
$b/v_1$,
and $c/v_1$;
likewise,
the matrix $\Gamma_2$ has elements $x/v_2$ and $y/v_2$.
In order to preserve the perturbation expansion,
we have required that,
for any particular solution in our fit,
all matrix elements of $\Gamma_1$ and $\Gamma_2$---and,
likewise,
of $\Delta_1$ and $\Delta_2$---do not exceed $4 \pi$ in modulus.

In the decays $B_d \rightarrow D^+ D^-$ one
has\footnote{Although there is
a loop-suppressed (but not CKM-suppressed)
penguin contribution to this decay,
this can be ignored due to the large experimental error.
It is only the sign of this observable
which will be of use below.}
\be
\lambda_{D^+ D^-} = \left( \frac{q}{p} \right)_{B_d}
\frac{V_{cb} V_{cd}^\ast}{V_{cb}^\ast V_{cd}}
\ee
and
\be
\sin{\left( 2 \beta \right)} = \mathrm{Im}\, \lambda_{D^+ D^-}.
\ee
We require $\sin{\left( 2 \beta \right)}$ computed in this way
to agree with the experimental value.
Notice that this path to $\sin{\left( 2 \beta \right)}$
does not include $M_{12}$ in the $K$ system.

The CP-violating phase
$\gamma = \arg{\left( - V_{ud} V_{cb} V_{ub}^\ast V_{cd}^\ast \right)}$
has been extracted from the decays $B^\pm \rightarrow D K^\pm$.
There are two experimentally allowed regions:
one region in which $\gamma \approx 70^\circ$
is in the first quadrant and has values consistent with the SM,
and another region with $\gamma \approx 70^\circ - 180^\circ$
in the third quadrant.
The solutions with $\gamma$ in the third quadrant, though,
are excluded by current measurements
of the semileptonic asymmetry in $B$ decays \cite{Uli}.
We have computed $\gamma$ in each of our fit points
and used it \textit{a posteriori} in our fit.

The experimental data discussed this far
potentially constrain the scalar masses $m_A$
and $m_\mathrm{eff}$ and the FCNC matrix $N_d$.
The most important constraints on $N_u$
come
from mixing (\textit{i.e.}\ $M_{12}$) in the $D$ system.
In the SM,
that mixing has three origins:
box diagrams,
dipenguin diagrams,
and long-distance physics.
The long-distance effects should be dominant
but are very difficult to estimate reliably.
Therefore,
we only require that the NP contribution by itself alone
should not exceed twice the experimental limit on $\Delta m_D$.

We want to comment on a set of points that
we have found at the first stage of our fit
and which display an inverted unitarity triangle,
\textit{i.e.}\ have a negative Jarlskog invariant \cite{Jarlskog}
$J_\mathrm{CKM} =
\mathrm{Im} \left( V_{us} V_{cb} V_{ub}^\ast V_{cs}^\ast \right)$.
Such points fit well the 15 observables used in the first stage,
but are all eliminated at the second stage of the fit,
because
they display $\gamma \approx - 70^\circ$,
in contradiction with experiment.
Besides,
some of the $J_\mathrm{CKM} < 0$ points suffer from the extra problem
that
they
rely on
dramatic contributions to $M_{12}$ in the $K$ system,
with $M_{12}^\mathrm{NP} \approx - 2 M_{12}^\mathrm{SM}$.
In these points the sign of $q/p$ in the $K$ system
is inverted  with respect to the SM.
As a result,
$\sin{\left( 2 \beta \right)}$ extracted from $\psi K_S$ decays
would have the opposite sign to the $\sin{\left( 2 \beta \right)}$
extracted from $D^+ D^-$ decays,
which is excluded by experiment.

\subsection{Two extra quantities}

CP violation has also been measured
in the decay $B_s \rightarrow \psi \phi$.
It is determined by
\be
\lambda_{\psi \phi} =
\left( \frac{q}{p} \right)_{B_s}
\frac{V_{cb} V_{cs}^\ast}{V_{cb}^\ast V_{cs}}.
\label{lambda-psiphi}
\ee
Using equation~(BLS-30.36),
we see that the SM leads to
$\lambda_{\psi \phi} = - \exp{\left( 2 i \beta_s \right)}$,
where $\beta_s$ is a phase in the CKM matrix which,
in the SM,
is of order a few percent.\footnote{In \cite {BLS}
the phase $\beta_s$ has been called $\epsilon$.}
Thus,
in the SM
\be
\sin{\left( 2 \beta_s \right)} = - \mathrm{Im}\, \lambda_{\psi \phi}.
\ee
We might have used
the current measurement of $\sin{\left( 2 \beta_s \right)}$
from the decays $B_s \rightarrow \psi \phi$
to constrain $M_{12}^\mathrm{NP}$ in the $B_s$ system.
However,
a recent average \cite{HFAG} excludes the SM at the $2.3\, \sigma$ level.
Our fits always yield a $\beta_s$ very close to its SM value;
thus,
our model does not provide a solution to
this discrepancy between the SM and experiment.

In this model,
direct CP violation is negligible in $D$ decays,
and therefore \cite{GNP}
\be
\arg{\left( \Gamma_{12}^\ast\, \frac{\bar A_{K^+K^-}}
{A_{K^+K^-}} \right)} = 0,
\ee
relates $\Gamma_{12}$ to the amplitudes for the decays
$D \rightarrow K^+ K^-$.
As a result,
\be
\phi_{12} = \arg{\left( M_{12} \Gamma_{12}^\ast \right)} =
- \arg{ \left( M_{12}^\ast\,
\frac{V_{cs} V_{us}^\ast}{V_{cs}^\ast V_{us}} \right)}.
\ee
As shown in reference \cite{GNP},
the theoretical parameter $\phi_{12}$
can be extracted from the experimental data.

\section{The fit: results}

After the two stages of our fit we still have many
points which have satisfied all the filtering criteria.
With those points
we have made a number of figures,
which we next present.

\begin{figure}[ht]
\epsfysize=6cm
\centerline{\epsfbox{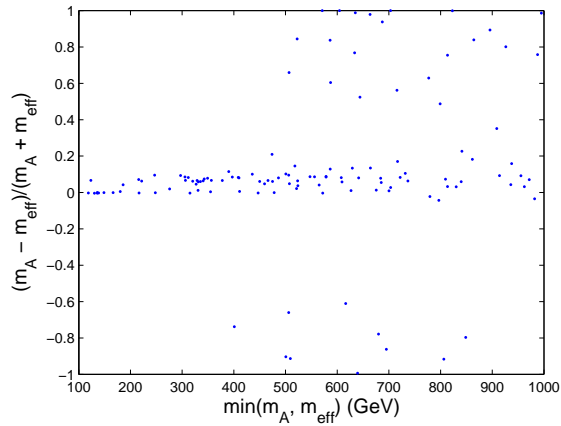}}
\caption{The masses of the scalars.
We do not display the points
where both $m_A$ and $m_\mathrm{eff}$ are larger than $1\, \mathrm{TeV}$.}
\label{fig:massdif}
\end{figure}
Figure~\ref{fig:massdif}
displays the asymmetry between $m_A$ and $m_\mathrm{eff}$
as a function of the smallest of those two masses.
Clearly,
if the scalar masses are both very large,
then the model is effectively like the SM,
except for the important fact that now the CKM
CP-violating phase does not arise
from complex hard (dimension-4) Yukawa couplings,
as in the SM,
but rather from a soft (dimension-2) CP-breaking term
in the scalar sector.
We find,
however,
that our model can have scalar masses
as small as a few hundred GeV,
especially when $m_A \approx m_\mathrm{eff}$.\footnote{Low scalar masses
may in some cases be excluded by other experimental constraints
that we have not taken into account,
for instance by top-quark decays.}
In this case of low scalar masses,
we have found that $M_{12}^\mathrm{NP} / M_{12}^\mathrm{SM}$
can be very large in the kaon sector,
but is not larger than 10\% in the $B_d$ and $B_s$ systems.

In order to quantify the latter statement,
we define \cite{Uli}
\be
M_{12} = M_{12}^\mathrm{SM} + M_{12}^\mathrm{NP}
= M_{12}^\mathrm{SM} \Delta,
\ee
where the SM limit corresponds to $\Delta = 1$.
We shall use a subscript $K, d, s$ in $\Delta$
to refer to the cases of the $K$ system,
$B_d$ system,
and $B_s$ system,
respectively.
The current measurements
do not agree well with the SM.
Setting $\Delta_K=1$ and excluding the measurement of $\beta_s$,
the CKMfitter Group \cite{CKMfitter} finds that
the current constraints on $\Delta_d$  and $\Delta_s$
exclude the SM at the $2.2\, \sigma$ and $1.9\, \sigma$ levels,
respectively.
The measurements of $\beta_s$ are much above the SM prediction
and further worsen this inconsistency \cite{Uli}.
Similar conclusions are drawn by the UTfit Collaboration \cite{UTfit}.

\begin{figure}[h]
\epsfysize=7cm
\centerline{\epsfbox{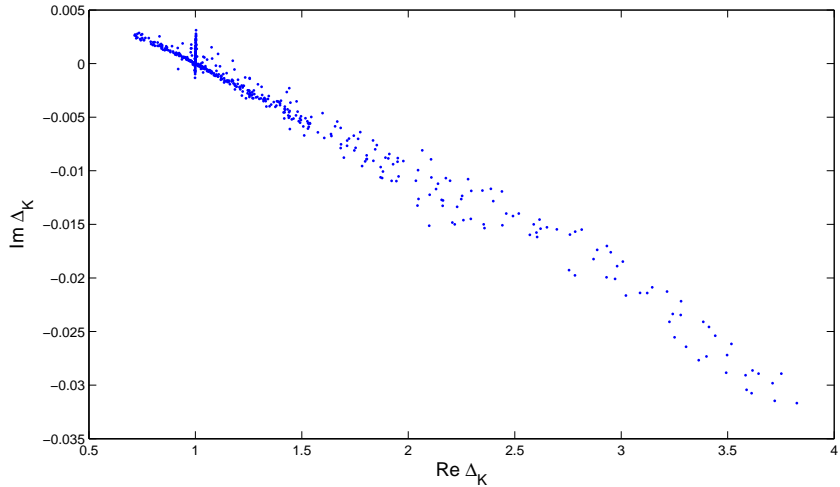}}
\caption{$\Delta$ parameter for the $K$ mesons.}
\label{fig:delK}
\end{figure}
\begin{figure}[h]
\epsfysize=7cm
\centerline{\epsfbox{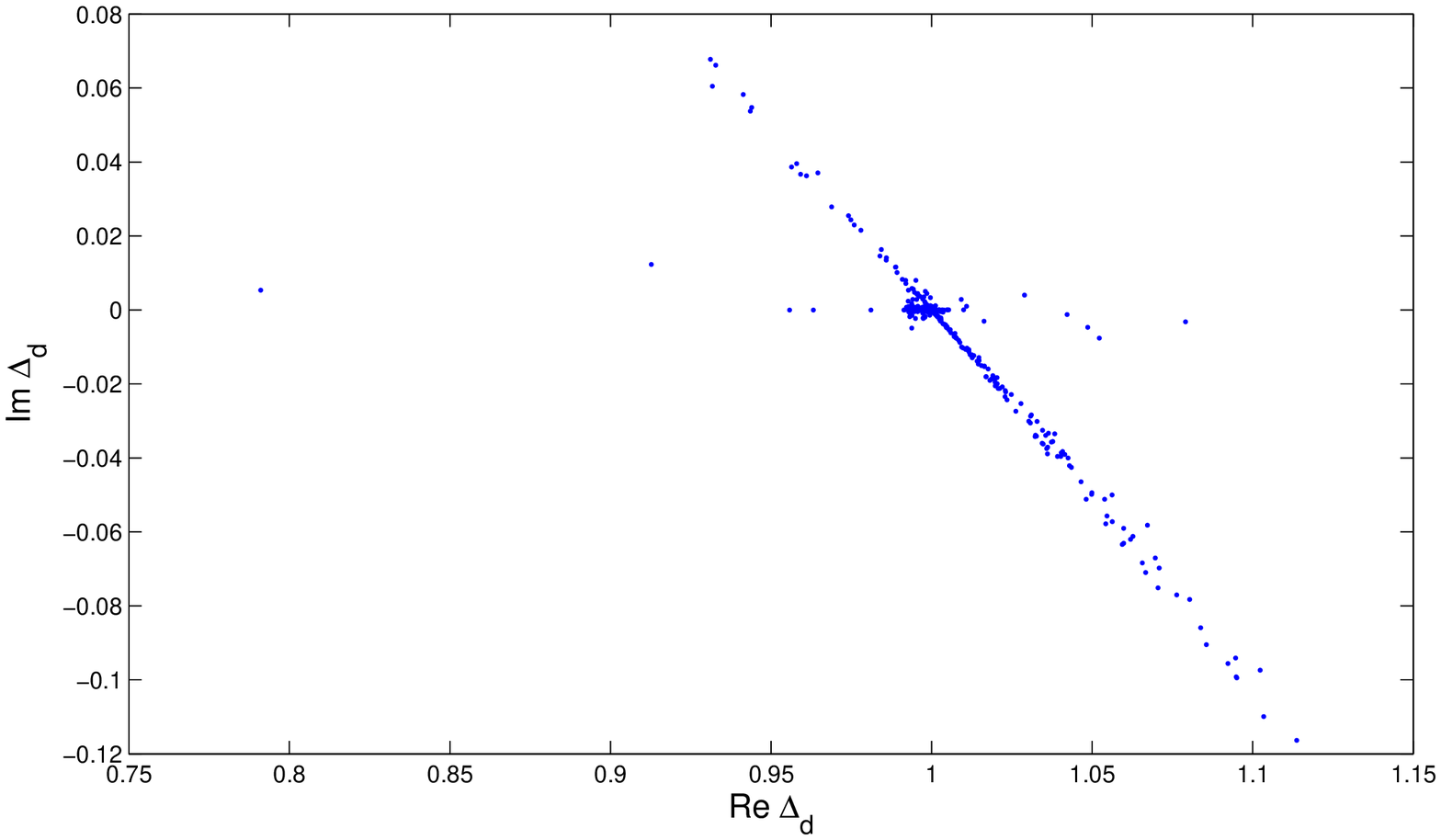}}
\caption{$\Delta$ parameter for the $B_d$ mesons.}
\label{fig:delBd}
\end{figure}
\begin{figure}[h]
\epsfysize=7cm
\centerline{\epsfbox{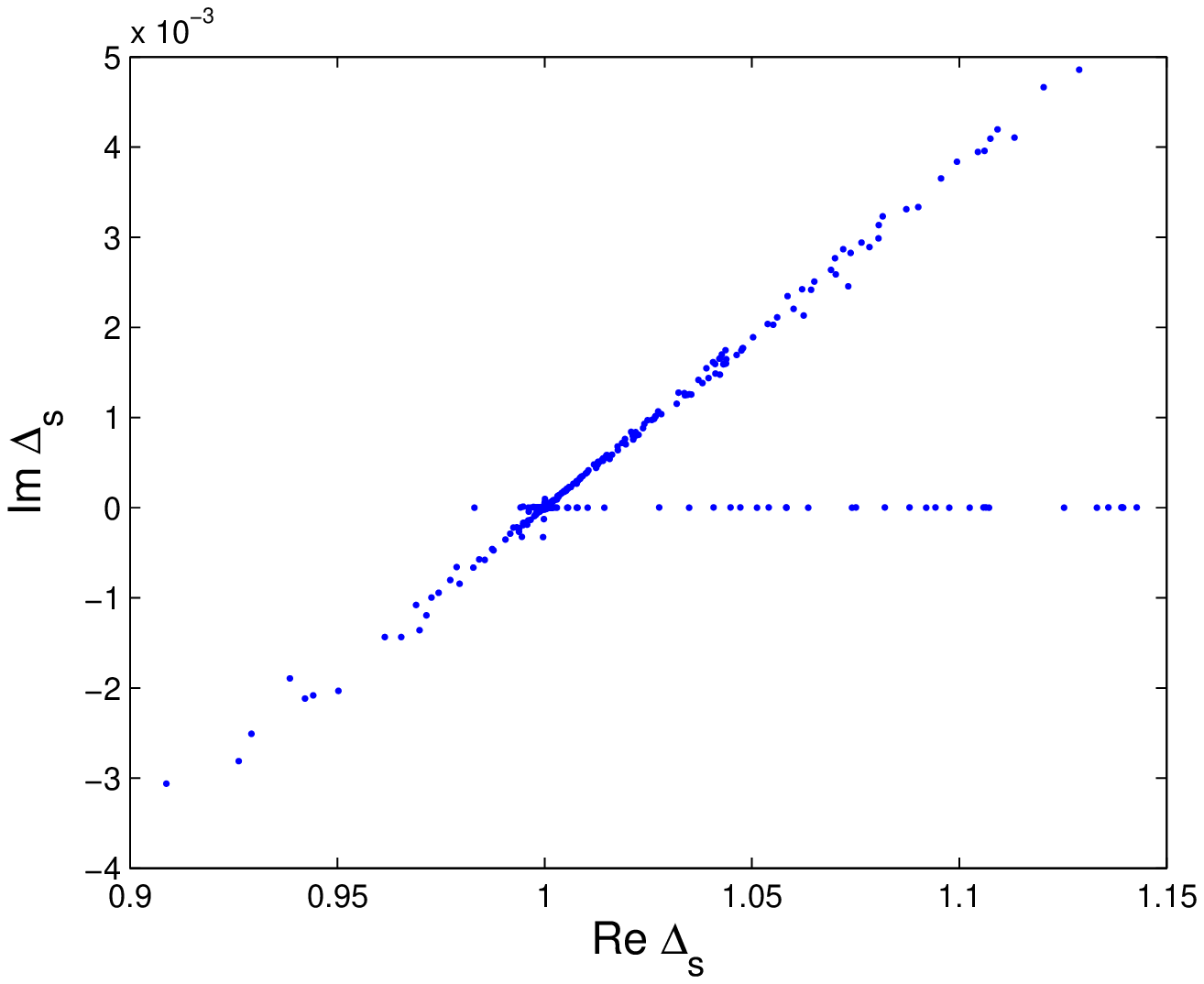}}
\caption{$\Delta$ parameter for the $B_s$ mesons.}
\label{fig:delBs}
\end{figure}
Figures~\ref{fig:delK},
\ref{fig:delBd},
and~\ref{fig:delBs} contain the results of our fits for $\Delta_K$,
$\Delta_d$,
and $\Delta_s$,
respectively.
We see that $\mathrm{Im}\, \Delta$ is in general quite small.
This is a reflection of the fact that in our model CP violation
lies exclusively in the CKM matrix while the matrix $N_d$ is real;
therefore $M_{12}^\mathrm{NP}$ is,
in our model,
real in all three neutral-meson systems.\footnote{We have neglected
potentially complex contributions to $M_{12}^\mathrm{NP}$ at loop level,
notably box diagrams involving intermediate charged scalars $C^\pm$.
This is consistent with our previously stated assumption that
the NP tree-level contributions to quark decays
are much smaller than the SM ones.}
We see in Figure~\ref{fig:delK}
that $\mathrm{Re}\, \Delta_K$ can be as large as three or four.
This freedom is due to the large uncertainty
in the long-distance contributions to $K$ mixing.
On the other hand,
since $\epsilon_K$ is small,
$\mathrm{Im}\, \Delta_K$ cannot be larger than two or three percent.
In the $B_d$ system,
changes of $\Delta_d$ of order 10\% relative to the SM
are possible both in the real and imaginary parts.
For some of our points this decreases slightly
the inconsistency of the SM with the experimental fits.
However,
this improvement is not dramatic because the experimental fits prefer
$\mathrm{Im}\, \Delta_d < 0$ and $\mathrm{Re}\, \Delta_d < 1$,
while our points with $\mathrm{Im}\, \Delta_d < 0$
have $\mathrm{Re}\, \Delta_d > 1$,
\textit{cf}.\ Figure~\ref{fig:delBd}.
In the $B_s$ system,
$\mathrm{Re}\, \Delta_s$ can differ from 1 by 10\% or so,
while $\mathrm{Im}\, \Delta_s$ remains at the 0.1\% level.
Thus,
in the $B_s$ system our model is as (in)consistent with experiment as the SM.

\begin{figure}[h]
\epsfysize=7cm
\centerline{\epsfbox{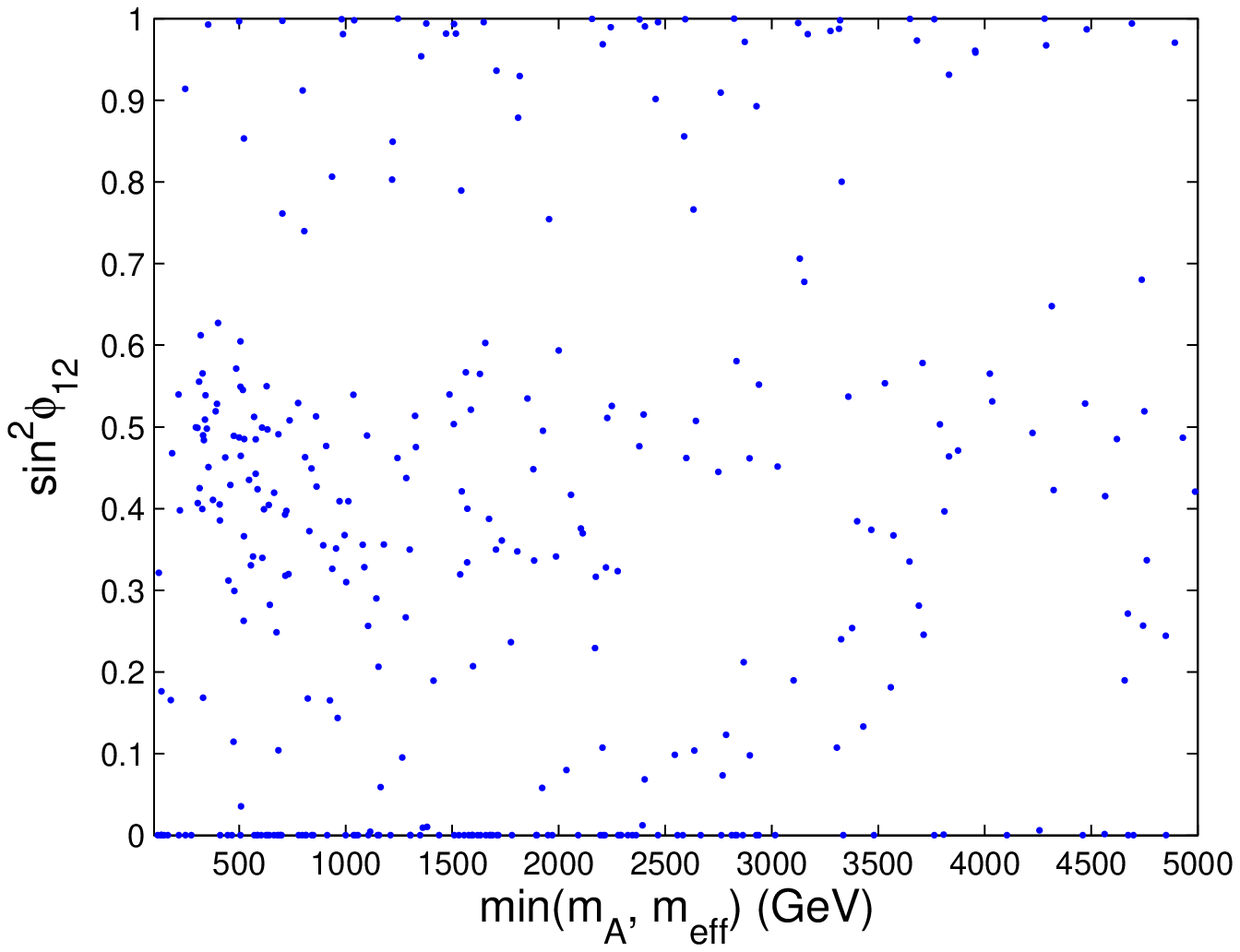}}
\caption{Predictions for $\sin^2{\phi_{12}}$.}
\label{fig:phi12}
\end{figure}
Figure~\ref{fig:phi12}
contains the predictions of our model for $\phi_{12}$,
based on the full set of our points
and using exclusively $M_{12}^\mathrm{NP}$,
\textit{i.e.}\ assuming $M_{12}^\mathrm{SM} = 0$.
We see in Figure~\ref{fig:phi12} that $\sin^2{\phi_{12}}$ is,
in our model,
arbitrary;
this illustrates how important CP violation in the $D$ system can be
in constraining models of new physics \cite{Petrov} such as ours.
Notice that the present experimental constraints on $\phi_{12}$
depend on a set of measurements which are highly correlated \cite{CKMfitter};
precise numbers are not available,
but we estimate, based on the method in \cite{GNP},
that $\sin^2{\phi_{12}} < 0.34$ at the $1\, \sigma$ level.

\section{Conclusions}

In this paper we have presented a two-Higgs-doublet model
with a $\mathbbm{Z}_3$ symmetry and the usual CP symmetry
in all the hard (dimension-four) terms but broken in one,
and only one,
soft (dimension-two) term in the scalar potential.
We have shown that this model displays a CP violation which,
just as in the SM,
is concentrated in the CKM matrix,
even though it has a completely different origin.
Contrary to most other 2HDMs,
our model exhibits CP violation neither in scalar--pseudoscalar mixing,
not in the scalar self-interactions,
nor in the matrices $N_{d,u}$ which parametrize
the flavour-changing Yukawa interactions of the neutral scalars.

Our model has only eleven parameters---ten moduli
and one phase---to fit the six quark masses
and the four independent observables of the CKM matrix.
When computing mixing in the neutral-meson--antimeson systems
one needs three extra parameters--- the ratio of VEVs,
the mass of the pseudoscalar,
and a weighted mass of the two scalars.
With these parameters one is able to fit most observables,
just as in the SM.
Remarkably,
many of these fits display scalar masses as low as $400\, \mathrm{GeV}$.

We have emphasized the relevance that a measurement of CP violation
in $D$-meson--antimeson mixing may have in eliminating some of our
fit points and, thus, in reducing the viable parameter space of our model.


\paragraph{Acknowledgements:}
The work of L.L.\ and of J.P.S.\ is funded by FCT through the projects
CERN/FP/109305/2009 and  U777-Plurianual,
and by the EU RTN project Marie Curie: MRTN-CT-2006-035505.
J.P.S.\ is grateful to Y.~Grossman,
Y.~Nir,
M.~Papucci,
and D.~Pirjol
for exchanges concerning solutions with negative $J_\textrm{CKM}$.

\appendix
\renewcommand{\theequation}{\Alph{section}\arabic{equation}}

\section{\label{sec:appendix} Input parameters}
\setcounter{equation}{0}

We have used in our fits
$G_F = 1.16639 \times 10^{-5}\, \mathrm{GeV}^{-2}$
and $m_W = 80.4\, \mathrm{GeV}$.
In the neutral-kaon system,
we have used $m_K = 497.614\, \mathrm{MeV}$,
$f_K = 155.5\, \mathrm{MeV}$;
for the QCD correction factors of equation~(BLS-17.16)
we have taken \cite{buchalla} $\eta_1 = 1.38$,
$\eta_2 = 0.57$,
and $\eta_3 = 0.47$.
For the ``bag parameter'' we have used ~\cite{CKMfitter} $B_K = 0.723$.
Our results do not depend crucially
on these inputs---small variations
thereof do not change our conclusions.

In the $B_d$ system,
we have used $m_{B_d} = 5.2795\, \mathrm{GeV}$,
$f_{B_d} = 190\, \mathrm{MeV}$,
$\eta_{B_d} = 0.55$,
and $B_{B_d} = 1.219$.
For the $B_s$ mesons,
\cite{CKMfitter,PDG},
$m_{B_s} = 5.366\, \mathrm{GeV}$,
$f_{B_s} = 228\, \mathrm{MeV}$,
$\eta_{B_s} = 0.55$,
and $B_{B_s} = 1.280$.
In the $D$ system,
$f_D = 232\, \mathrm{MeV}$~\cite{lin}
and $m_D = 1.86483\, \mathrm{GeV}$~\cite{PDG}.

We next present two of our fit points:
one with low masses $m_A$ and $m_\mathrm{eff}$
and another one in which one of the masses is low
and the other one much larger.
\begin{itemize}
\item
First point: $a = 105\, \mathrm{MeV}$,
$b = 15.2\, \mathrm{MeV}$,
$c = 6\, \mathrm{MeV}$,
$x = 4.1387\, \mathrm{GeV}$,
$y = 24.2\, \mathrm{MeV}$,
$a' = 169.6184\, \mathrm{GeV}$,
$b' = -8.5\, \mathrm{MeV}$,
$c' = 1.2823\, \mathrm{GeV}$,
$x' = 6.8821\, \mathrm{GeV}$,
$y' = -1.8\, \mathrm{MeV}$,
$\theta = 5.6548\, \mathrm{rad}$,
$r = 0.4277$,
$m_A = 415.6327\, \mathrm{GeV}$,
and $m_\mathrm{eff} = 411.0434\, \mathrm{GeV}$.
\item
Second point: $a = 17.5\, \mathrm{MeV}$,
$b = 180.7\, \mathrm{MeV}$,
$c = -27.8\, \mathrm{MeV}$,
$x = 4.2504\, \mathrm{GeV}$,
$y = 74.9\, \mathrm{MeV}$,
$a' = 172.8953\, \mathrm{GeV}$,
$b' = -29.6\, \mathrm{MeV}$,
$c' = -2.7\, \mathrm{MeV}$,
$x' = -600.3\, \mathrm{MeV}$,
$y' = 1.2920\, \mathrm{GeV}$,
$\theta = 3.6987\, \mathrm{rad}$,
$r = 0.8217$,
$m_A = 400.9344\, \mathrm{GeV}$,
and $m_\mathrm{eff} = 2.6596688\, \mathrm{TeV}$.
\end{itemize}

With these inputs one obtains the following values for the observables:
\begin{itemize}
\item First point:
$m_d = 5.8\, \mathrm{MeV}$,
$m_s = 107.8\, \mathrm{MeV}$,
$m_b = 4.1388\, \mathrm{GeV}$,
$m_u = 1.8\, \mathrm{MeV}$,
$m_c = 1.2813\, \mathrm{GeV}$,
$m_t = 169.758\, \mathrm{GeV}$,
$\left| V_{us} \right| = 0.2256$,
$\left| V_{cb} \right| = 0.0405$,
$\left| V_{ub} \right| = 0.0037$,
$\left| V_{td} \right| = 0.0087$,
$\Delta m_K = 2.93 \times 10^{-6}\, \mathrm{eV}$,
$\Delta m_{B_d} = 3.55 \times 10^{-4}\, \mathrm{eV}$,
$\Delta m_{B_s} = 1.13 \times 10^{-2}\, \mathrm{eV}$,
$\left| \epsilon_K \right| = 2.227 \times 10^{-3}$,
$J_\textrm{CKM} = 3.113 \times 10^{-5}$,
$\sin{\left( 2 \alpha \right)} = 0.1714$,
$\sin{\left( 2 \beta \right)} = 0.7128$,\footnote{This is
the $\sin{\left( 2 \beta \right)}$
which is obtained from the decays $B_d \rightarrow \psi K_S$.
The value obtained from $B_d \rightarrow D^+ D^-$ is 0.7189.}
$\sin{\left( 2 \beta_s \right)} = -0.0397$,
$\gamma = 71.96^\circ$,
$\Delta_K = 1.3242 - 0.0032\, i$,
$\Delta_d = 0.9996 + 7.2444 \times 10^{-5}\, i$,
$\Delta_s = 1.0000 - 9.6084 \times 10^{-7}\, i$.
\item Second point:
$m_d = 6.0\, \mathrm{MeV}$,
$m_s = 81.5\, \mathrm{MeV}$,
$m_b = 4.2542\, \mathrm{GeV}$,
$m_u = 2.7\, \mathrm{MeV}$,
$m_c = 1.2923\, \mathrm{GeV}$,
$m_t = 172.8963\, \mathrm{GeV}$,
$\left| V_{us} \right| = 0.2249$,
$\left| V_{cb} \right| = 0.0425$,
$\left| V_{ub} \right| = 0.0037$,
$\left| V_{td} \right| = 0.0089$,
$\Delta m_K = 4.98 \times 10^{-6}\, \mathrm{eV}$,
$\Delta m_{B_d} = 3.57 \times 10^{-4}\, \mathrm{eV}$,
$\Delta m_{B_s} = 1.16 \times 10^{-2}\, \mathrm{eV}$,
$\left| \epsilon_K \right| = 2.128 \times 10^{-3}$,
$J_\textrm{CKM} = 3.211 \times 10^{-5}$,
$\sin{\left( 2 \alpha \right)} = 0.1167$,
$\sin{\left( 2 \beta \right)} = 0.7450$,\footnote{The
$\sin{\left( 2 \beta \right)}$ obtained from
$B_d \rightarrow D^+ D^-$ is 0.7487.}
$\sin{\left( 2 \beta_s \right)} = -0.0407$,
$\gamma = 69.11^\circ$,
$\Delta_K = 2.2183 - 0.0150\, i$,
$\Delta_d = 0.9311 + 0.0678\, i$,
$\Delta_s = 0.9088 - 0.0031\, i$.
\end{itemize}

\section{Oblique parameters} \label{app:oblique}
\setcounter{equation}{0}

Relevant contraints on the scalar spectrum
of a two-Higgs-doublet model arise from consideration
of the so-called `oblique parameters',
especially of the parameters $S$ and $T$.\footnote{The other
oblique parameters are usually very small and,
therefore,
irrelevant.
We have checked this explicitly for some of our points.}
Formulae for those parameters in a general MHDM
have been presented in ref.~\cite{ogreid}.
In our particular 2HDM,
one has
\ba
T &=& \frac{1}{16 \pi s_w^2 m_W^2} \left[
\cos^2{\psi}\, f (m_C^2, m_1^2)
+ \sin^2{\psi}\, f (m_C^2, m_2^2) + f (m_C^2, m_A^2)
\right. \no & &
- \sin^2{\psi}\, f (m_2^2, m_A^2) - \cos^2{\psi}\, f (m_1^2, m_A^2)
\no & &
\left. + \sin^2{\psi}\, f^\prime (m_1^2)
+ \cos^2{\psi}\, f^\prime (m_2^2) - f^\prime (m_H^2)
\right],
\label{T}
\ea
where
\be
f (x, y) = \left\{ \begin{array}{lcl}
{\displaystyle \frac{x + y}{2} - \frac{x y}{x - y}\, \ln{\frac{x}{y}}}
& \Leftarrow & x \neq y,
\\
0 & \Leftarrow & x = y,
\end{array} \right.
\ee
\be
f^\prime (m^2) = 3 \left[ f (m_Z^2, m^2) - f (m_W^2, m^2) \right].
\label{fprime}
\ee
In equations~(\ref{T}) and~(\ref{fprime}),
$m_C$ is the mass of the charged scalars $C^\pm$,
$m_W$ is the mass of the $W^\pm$,
$m_Z$ is the mass of the $Z^0$,
$m_H$ is the mass of the SM Higgs particle,
and $s_w^2 = 1 - m_W^2 / m_Z^2$.
The expression for $S$ is
\ba
S &=& \frac{1}{24 \pi} \left[
(1 - 2 s_w^2)^2\, g (x_C, x_C)
+ \sin^2{\psi}\, g (x_2, x_A) + \cos^2{\psi}\, g (x_1, x_A)
\right. \no & & \left.
+ \sin^2{\psi}\, \hat g (x_1)
+ \cos^2{\psi}\, \hat g (x_2)
- \hat g (x_H)
+ \ln{\frac{m_1^2 m_2^2 m_A^2}{m_C^4 m_H^2}} \right],
\label{S}
\ea
where $x_k = m_K^2 / m_Z^2$ for $k = 1, 2, A, C$.
The functions $g (x, y)$ and $\hat g (x)$ in equation~(\ref{S})
are in the second paper of ref.~\cite{ogreid}.\footnote{One has
$g (x, y) \equiv G (xz, yz, z)$ and $\hat g (x) \equiv \hat G (xz, z)$,
with the functions $G ( I, J, Q)$ in equation~(C2)
and $\hat G (I, Q)$ in equation~(C5) of ref.~\cite{ogreid}.}

For each of our fit points we only have
the masses $m_\mathrm{eff}$---in eq.~(\ref{eq:meff})---and $m_A$.
Equation~(\ref{eq:meff}) may be solved for the mixing angle $\psi$,
yielding
\be
\sin^2{\psi} = \frac{m_1^2}{m_\mathrm{eff}^2}\,
\frac{m_2^2 - m_\mathrm{eff}^2}{m_2^2 - m_1^2},
\quad
\cos^2{\psi} = \frac{m_2^2}{m_\mathrm{eff}^2}\,
\frac{m_1^2 - m_\mathrm{eff}^2}{m_1^2 - m_2^2}.
\label{psi}
\ee
Therefore,
either $m_1 \le m_\mathrm{eff} \le m_2$ or $m_2 \le m_\mathrm{eff} \le m_1$.

In order to check whether each of our fit points---defined
by given values of $m_\mathrm{eff}$ and $m_A$---is compatible
with the experimental bounds on the oblique parameters~\cite{erler},
we have inputted various values of $m_{1,2}$ and $m_C$.
From $m_1$ and $m_2$ we have computed $\psi$ through equation~(\ref{psi})
and then the oblique parameters $S$ and $T$.
With a fast fitting program we have been
able to find,
for a large part of our fit points,
values of $m_1$,
$m_2$,
and $m_C$ such that $S$ and $T$ result compatible
with the experimental bounds.\footnote{We have used $m_H = 117\, \mathrm{GeV}$
in accordance with one of the experimental ellipses
in Figure~10.4 of ref~\cite{erler}.}
The masses $m_{1,2,C}$ can be chosen such that the parameter~$T$
does not result too large (either positive or negative).
The parameter~$S$ usually turns out to be positive
and relatively large
($S\, \raise.3ex\hbox{$>$\kern-.75em\lower1ex\hbox{$\sim$}}\, 0.1$),
but for most\footnote{We have \emph{not} been able to explicitly find out,
for \emph{all} of our fit points,
values of $m_{1,2,C}$ such that both $S$ and $T$ agree with
the experimental bounds,
but we cannot exclude that that is possible.}
of our points it can be made compatible with the experimental bounds.

As an example,
one of our fit points has $m_\mathrm{eff} = 5.4711\, \mathrm{TeV}$
and $m_A = 679.9875\, \mathrm{GeV}$.
Choosing $m_1 = 0.99\, m_\mathrm{eff}$,
$m_2 = 2.0283\, m_\mathrm{eff}$,
and $m_C = 2\, m_\mathrm{eff}$,
one obtains $S = 0.21$ and $T = 0.19$.

\section{Direct LEP bounds}
\setcounter{equation}{0}

In Figure~\ref{fig:massdif} we have shown that our fit
sometimes yields scalar masses as low as 100$\,$GeV.
Since the current limit from LEP is 114.4$\,$GeV~\cite{PDG},
it is necessary to verify that our results
do not contradict that bound.
The LEP result is obtained by looking at the
associated production of a scalar particle and a $Z^0$ boson,
$e^+e^- \rightarrow Z h$,
which is possible due to the vertex $ZZh$.
For a 2HDM there is also $ZH$ production,
but not $ZA$ production,
since there is no $ZZA$ vertex.
Moreover,
as compared to the SM,
the coupling of the vertex $ZZh$ ($ZZH$)
is reduced by factors related to the mixing angle $\psi$
in equation~\eqref{eq:meff}.
Indeed,
in our model one has
\be
\frac
{\sigma^{\rm 2HDM} \left( e^+ e^- \rightarrow ZS \right)}
{\sigma^{\rm SM} \left( e^+ e^- \rightarrow ZS \right)}
= g^2_{ZZS},
\ee
where $g^2_{ZZS} = \sin^2{\psi}$
($\cos^2{\psi}$)
if $S=h$
($S=H$).
Therefore,
in a 2HDM it is possible to have scalars with masses
lower than the LEP bound,
provided those scalars couple more weakly to $ZZ$ than in the SM.

\begin{figure}
\epsfysize=7cm
\centerline{\epsfbox{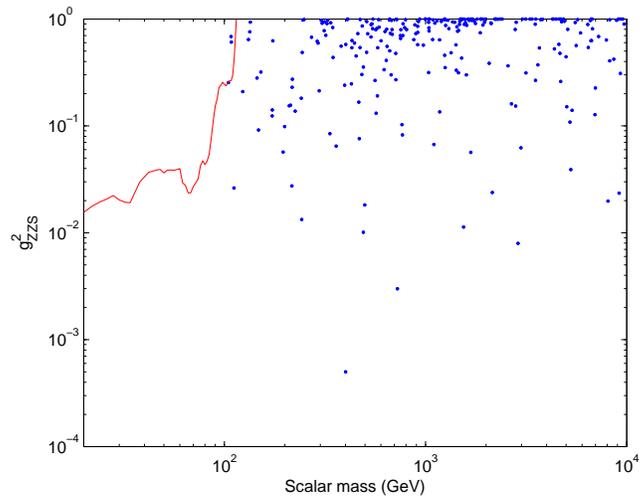}}
\caption{Constraints on scalar masses from direct searches at LEP
(data taken from ref.~\cite{lep}).}
\label{fig:lep}
\end{figure}

As explained before,
our fit to the quark masses,
CKM matrix elements,
and CP-violating quantities
has produced a large number of acceptable points in parameter space.
Out of those,
as seen in Appendix~B,
the vast majority conforms to the existing constraints
on the oblique parameters.
In Figure~\ref{fig:lep} we plot,
for $h$ and $H$ simultaneously,
the comparison between
the set of points which have passed the oblique-parameter fit
and the experimental data from the direct searches at LEP;
acceptable points must be below and to the right of the solid line in the plot.
We see that,
with the exception of only three points,
the parameter space that we have found agrees perfectly with the LEP data.
(As with the fit to the oblique parameters,
we cannot exclude hat other values of $m_{1,2,C}$ can be found,
such that all the points agree with the LEP experimental bounds.)

We have also looked at the existing LEP bounds
on scalar--pseudoscalar production.
Those bounds extend to 225$\,$GeV
in the sum of the masses of the scalar and the pseudoscalar.
We have found that all our points
which survive the LEP bounds on $Z^0$--scalar production
display a sum of the masses of the scalar and the pseudoscalar
which exceeds 225$\,$GeV.
Therefore,
all those points also survive the LEP bounds
on scalar--pseudocalar production.

\end{document}